\documentclass[aps,prc,twocolumn,showpacs,superscriptaddress,floatfix,10pt]{revtex4}
\usepackage{bm}
\usepackage{amssymb}
\usepackage{amsmath}
\usepackage{graphicx}
\begin{document}
\title{Microscopic sub-barrier fusion calculations for the neutron star crust}

\author{A.S. Umar}\email{umar@compsci.cas.vanderbilt.edu}
\author{V.E. Oberacker}\email{volker.e.oberacker@vanderbilt.edu}
\affiliation{Department of Physics and Astronomy, Vanderbilt University, Nashville, TN 37235, USA}
\author{C. J. Horowitz}\email{horowit@indiana.edu}
\affiliation{Department of Physics and CEEM,
             Indiana University, Bloomington, IN 47405, USA}
\affiliation{Department of Physics and Astronomy, University of Tennessee, Knoxville, Tennessee 37996, USA}
\affiliation{Physics Division, Oak Ridge National Laboratory, P.O. Box 2008, Oak Ridge, TN 37831, USA}
\date{\today}
\begin{abstract}
Fusion of very neutron rich nuclei may be important to determine the composition and heating of the crust of accreting
neutron stars. Fusion cross sections are calculated using time-dependent Hartree-Fock theory coupled with
density-constrained Hartree-Fock calculations to deduce an effective potential. Systems studied include $^{16}$O+$^{16}$O,
$^{16}$O+$^{24}$O, $^{24}$O+$^{24}$O, $^{12}$C+$^{16}$O, and $^{12}$C+$^{24}$O. We find remarkable agreement with
experimental cross sections for the fusion of stable nuclei. Our simulations use the SLy4 Skyrme force that has been
previously fit to the properties of stable nuclei, and no parameters have been fit to fusion data. We compare our
results to the simple S\~{a}o Paulo static barrier penetration model. For the asymmetric systems $^{12}$C+$^{24}$O or
$^{16}$O+$^{24}$O we predict an order of magnitude larger cross section than those predicted by the S\~{a}o Paulo model.
This is likely due to the transfer of neutrons from the very neutron rich nucleus to the stable nucleus and
dynamical rearrangements of the nuclear densities during the collision process. These effects are not included in
potential models. This enhancement of
fusion cross sections, for very neutron rich nuclei, can be tested in the laboratory with radioactive beams.
\end{abstract}
\smallskip
\maketitle
\raggedbottom
\section{Introduction}
There is broad interest in the sub-barrier fusion of neutron rich nuclei, in astrophysics as we discuss in Sec.
\ref{sec.astro}, in the laboratory as discussed in Sec.~\ref{sec.lab}, and in theory, see Sec.~\ref{sec.theory}.

\subsection{Fusion in accreting neutron stars}
\label{sec.astro}
Neutron stars, in binary systems, can accrete material from their companions. This material undergoes a variety of
nuclear reactions.  First at low densities, conventional thermonuclear fusion takes place, see for example~\cite{xrayburstreview,rpash}.
Next at higher densities, the rising electron Fermi energy induces a series of electron
captures~\cite{gupta} to produce increasingly neutron rich nuclei. Finally at high densities, these very neutron rich
nuclei can fuse via pycnonuclear reactions. Pycnonuclear fusion is induced by quantum zero point
motion~\cite{pycno}. The energy released, and the densities at which these reactions occur, are important for
determining the temperature and composition profile of accreting neutron star crusts.

Superbursts are very energetic X-ray bursts from accreting neutron stars that are thought to involve the unstable
thermonuclear burning of carbon~\cite{superbursts, superbursts2}. However, the best current simulations do not
reproduce the conditions needed for carbon ignition because they have too low temperatures~\cite{superignition}.  An
additional heat source, from fusion or other reactions, could raise the temperature and allow carbon ignition at
densities that reproduce observed burst frequencies.  Alternatively, there could be a very low energy resonance in the
$^{12}$C -$^{12}$C fusion cross-section that could explain superburst ignition~\cite{c12cooper}.

Haensel and Zdunik have calculated pycnonuclear fusion reactions at great densities in the inner crust of neutron stars
\cite{haensel}  using a very simple crust composition consisting of a single average nucleus.  See also the work by
Sato\cite{sato}.  Gupta et al. mention the potential impact of electron capture and neutron emission cascades in their
model on fusion~\cite{guptacascades}. Here a cascade of multiple electron captures and neutron emissions can reduce the
atomic number of nuclei until the lower Coulomb barrier allows rapid pycnonuclear fusion.

Horowitz et al.~\cite{horowitzfusion} calculate the enhancement in fusion rates from strong ion screening using molecular
dynamics simulations, and find that $^{24}$O + $^{24}$O can fuse near $10^{11}$~g/cm$^3$, just before neutron drip.  Lau
uses a reaction network to follow the composition of an accreting fluid element in the neutron star crust
\cite{rita}.  She considers a range of fusion reactions involving neutron rich isotopes of chemical elements that
include carbon to magnesium~\cite{rita2}. These reactions tend to occur at densities just above neutron drip ($\geq$
few $\times 10^{11}$~g/cm$^3$) and beyond. Therefore we expect fusion reactions of neutron rich isotopes, near the drip
line, to be important.  Furthermore, this fusion can take place in the background neutron gas that is present in the
inner crust of a neutron star.

\subsection{Fusion experiments with neutron rich radioactive beams}
\label{sec.lab}
Recent advances in radioactive beam technologies have opened up new experimental possibilities to study fusion of
neutron rich nuclei. Furthermore, near barrier fusion cross sections are relatively large so experiments are feasible
with modest beam intensities.  For example, at the GANIL-SPIRAL facility a reaccelerated beam of $^{20}$O was used to
measure near- and sub-barrier fusion of $^{20}$O on $^{12}$C~\cite{desouza}. In addition, measurements are possible at
the TRIUMF ISAC facility and in the near future at the NSCL ReA3-6 reaccelerated beam facility.  Note that the dynamics
of the neutron rich skin of these nuclei can enhance the cross-section over that predicted by a simple static barrier
penetration model. For example, neutrons may be transferred from the neutron rich beam to the stable target. Fusion of
very neutron rich nuclei, near the drip line, raise very interesting nuclear structure and nuclear dynamics questions.

\subsection{Calculations of sub-barrier fusion cross sections}
\label{sec.theory}
In the absence of a practical ab-initio quantal many-body theory for sub-barrier fusion all approaches involve
the calculation of an ion-ion potential barrier, usually as a function of the nuclear separation coordinate $R$,
and the solution of the corresponding one-body Schr\"{o}dinger equation for the transmission probability and
the fusion cross-sections.
Many of the phenomenological and semi-microscopic potentials for fusion utilize the double-folding method~\cite{SL79,RO83a},
which is based on the physical assumption of {\it frozen densities} or the {\it sudden}
approximation. As the name suggests, in this approximation the nuclear densities
are unchanged during the computation of the ion-ion potential as a function
of the internuclear distance.
While asymptotically fusion potentials may be determined from
Coulomb and centrifugal interactions, the short distance
behavior strongly depends on the nuclear surface properties
and the readjustments of the combined nuclear system,
resulting in potential pockets, which determine the
characteristics of the compound nuclear system.
For this reason approaches based on frozen densities start to fail as the nuclear surface overlap
begins to be substantial due to lack of rearrangements and exchange effects.
This is the situation for fusion at deep sub-barrier energies for which the inner turning point
of the potential barrier happens at a large density overlap.
In the coupled-channel approach this has been addressed with the addition of a repulsive core
potential at small nuclear separations~\cite{Esb06,Esb08}.

During the past several years, we have developed a new microscopic approach for
calculating heavy-ion interaction potentials that incorporates
all of the dynamical entrance channel effects included in the
time-dependent Hartree-Fock (TDHF) description of the collision
process~\cite{UO06a}. The method is based on the TDHF evolution of
the nuclear system coupled with density-constrained
Hartree-Fock calculations (DC-TDHF) to obtain the ion-ion interaction potential.
The formalism was applied to study fusion cross-sections for
the systems $^{132}$Sn+$^{64}$Ni~\cite{UO07a}, $^{64}$Ni+$^{64}$Ni~\cite{UO08a},
$^{16}$O+$^{208}$Pb~\cite{UO09b}, $^{132,124}$Sn+$^{96}$Zr~\cite{OU10b}, and to reactions leading to
the superheavy element $Z=112$~\cite{UO10a}, among others. In all cases a
good agreement between the measured fusion cross-sections and the DC-TDHF results were found.
This is rather remarkable given the fact that the only input in DC-TDHF is the
nuclear effective interaction, and there are no adjustable parameters.

There is a great deal of experimental information on low energy fusion cross-sections for light stable nuclei such as
$^{12}$C~\cite{c12o16expdata_1,c12o16expdata_2} and $^{16}$O~\cite{o16o16expdata_1,o16o16expdata_2}.  For these nuclei,
barrier penetration models work well for energies near the Coulomb barrier~\cite{bp}.
However, recently Jiang et al. discussed fusion hindrance at extreme sub coulomb barrier energies
\cite{c12o16expdata_3}. Much less information is available for the fusion of very neutron rich light nuclei.

The S\~{a}o Paulo model of fusion calculates an effective potential based on the density overlap between colliding nuclei
\cite{sp}.  Sub-barrier fusion cross-sections can then be calculated via tunneling.  The model can be easily applied to
a very large range of fusion reactions and qualitatively reproduces many experimental cross-sections~\cite{gasques07}.
Recently this model was used to tabulate astrophysical $S$ factors describing fusion of many carbon, oxygen, neon and
magnesium isotopes for use in astrophysical simulations~\cite{M.Beard}, see also Ref.~\cite{M.Beard2}.

This paper is organized as follows.  Our time dependent Hartree Fock formalism is discussed in Sec.~\ref{sec.formalism},
while Sec.~\ref{sec.results} presents results for effective potentials and fusion cross sections.  Finally, these
results are discussed and we conclude in Sec.~\ref{sec.conclusions}.

\section{Formalism}
\label{sec.formalism}

We describe our time-dependent Hartree-Fock based formalism in Sec.~\ref{subsec.TDHF} and then, for comparison, review the
simple S\~{a}o Paulo barrier penetration model in Sec.~\ref{subsec.saopaulo}.

\subsection{TDHF and DC-TDHF}
\label{subsec.TDHF}

In this paper, we utilize the DC-TDHF method for the calculation of the ion-ion potential barriers.
It is generally acknowledged that the TDHF method provides a
useful foundation for a fully microscopic many-body theory of low-energy heavy-ion reactions
\cite{Ne82} based on a time-dependent nuclear energy-density functional. During 1970's and 1980's the TDHF
theory has been widely used in the study of fusion excitation functions,
deep-inelastic scattering of heavy mass systems, and nuclear molecular resonances,
while providing a natural foundation for many other studies. An
account of some of the earlier TDHF applications can be found in Refs.~\cite{Ne82,DDKS}.
With modern supercomputers it has become feasible to carry out very accurate nuclear
structure and reaction studies in full three-dimensional space and using the complete
form of the nuclear energy-density functional.
The DC-TDHF approach is analogous to calculating microscopic potential
energy surfaces with the constrained Hartree-Fock method using e.g.
quadrupole and octupole constraints. However, in this approach
the TDHF time-evolution takes place with no restrictions.
At certain times during the time evolution we
perform a parallel static Hartree-Fock minimization while holding the neutron and proton densities constrained
to be the corresponding instantaneous TDHF densities.
In essence, this gives us the
TDHF dynamical path in relation to the multi-dimensional static energy surface
of the combined nuclear system.
The DC-TDHF approach~\cite{UO06a} provides the means for extracting ion-ion potentials from
the TDHF evolution of the nuclear collision as follows:
The ion-ion interaction potential is given by
\begin{equation}
V_\mathrm{DC}(\bar{R})=E_{\mathrm{DC}}(\bar{R})-E_{\mathrm{A_{1}}}-E_{\mathrm{A_{2}}}\;,
\label{eq:vr}
\end{equation}
where $E_{\mathrm{DC}}$ is the density-constrained energy at the instantaneous
separation between nuclear centers $\bar{R}(t)$, while $E_{\mathrm{A_{1}}}$ and $E_{\mathrm{A_{2}}}$ are the binding energies of
the two nuclei obtained with the same energy-density functional.
The TDHF evolution also provides us with a coordinate-dependent mass, $M(\bar{R})$, which should be used in
solving the resulting Schr\"{o}dinger equation, using the conservation of energy
\begin{equation}
M(\bar{R})=\frac{2[E_{\mathrm{c.m.}}-V_{DC}(\bar{R})]}{\dot{\bar{R}}^{2}}\;,
\label{eq:mr}
\end{equation}
where the collective velocity $\dot{\bar{R}}$ is directly obtained from the TDHF evolution.
The $\bar{R}$-dependence of this mass at lower energies is
very similar to the one found in constrained Hartree-Fock calculations~\cite{GRR83} with a constraint
on the quadrupole moment.
On the other hand, at higher energies the coordinate dependent mass essentially becomes flat,
which is again a sign that most dynamical effects are contained at lower energies.
The peak at small $\bar{R}$ values is
due to the fact that the center-of-mass energy is above the barrier and the
denominator of Eq.~(\ref{eq:mr}) becomes small due to the slowdown of the ions.

All of the dynamical features included in TDHF are naturally included in the DC-TDHF potentials.
These effects include neck formation, mass exchange,
internal excitations, deformation effects to all order, among others.
Alternatively, instead of solving the Schr\"odinger equation with coordinate dependent
mass parameter $M(\bar{R})$ for the heavy-ion potential $V_{\mathrm{DC}}(\bar{R})$, we can instead use the constant
reduced mass $\mu$ and transfer the coordinate-dependence of the mass to a scaled
potential $V(R)$ using the well known exact point transformation~\cite{GRR83,UO09b}
\begin{equation}
dR=\left(\frac{M(\bar{R})}{\mu}\right)^{\frac{1}{2}}d\bar{R}\;.
\label{eq:mrbar}
\end{equation}
The potential $V(R)$, which includes the coordinate-dependent mass effects differs from the
$V_{\mathrm{DC}}(\bar{R})$ only in the interior region of the barrier. Further details can
be found in Ref.~\cite{UO09b}.
\begin{figure}[!htb]
\includegraphics*[width=6.6cm]{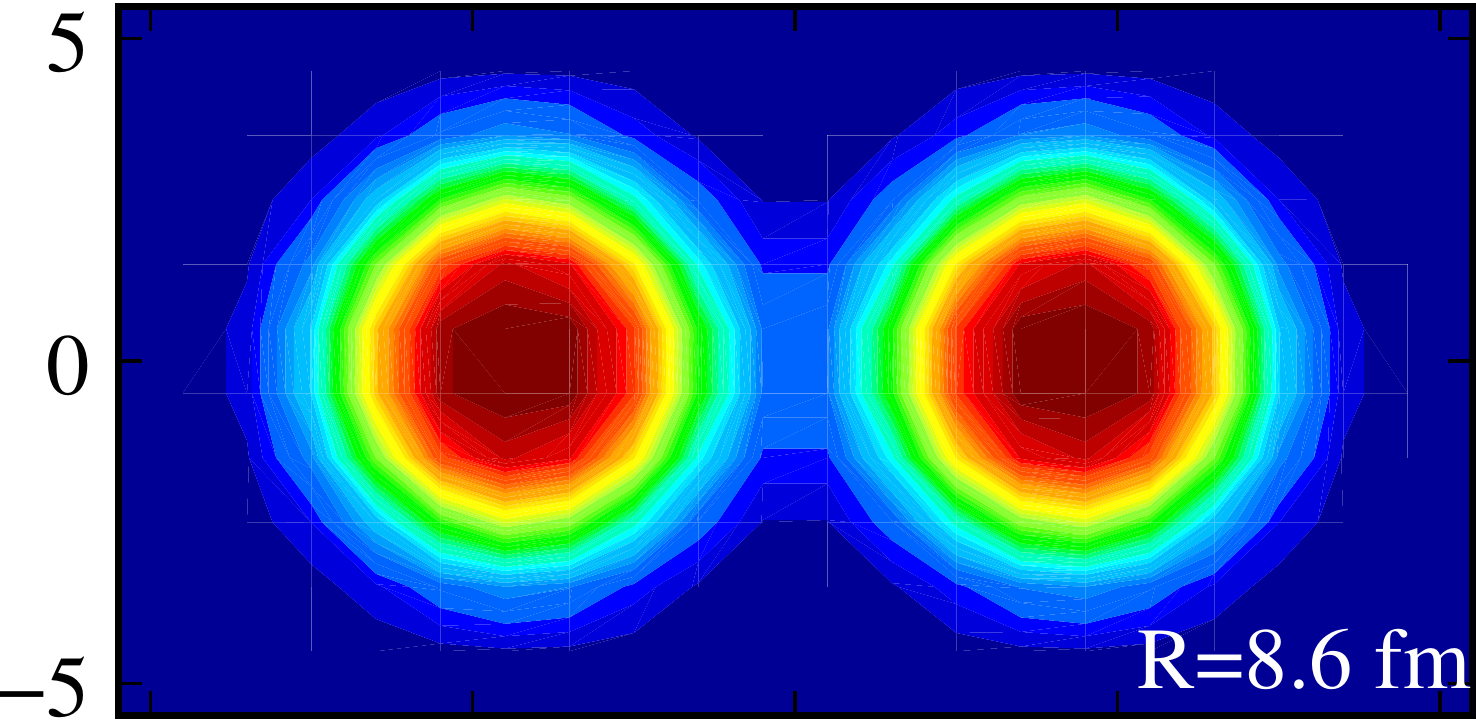}\vspace{-2pt}
\includegraphics*[width=6.6cm]{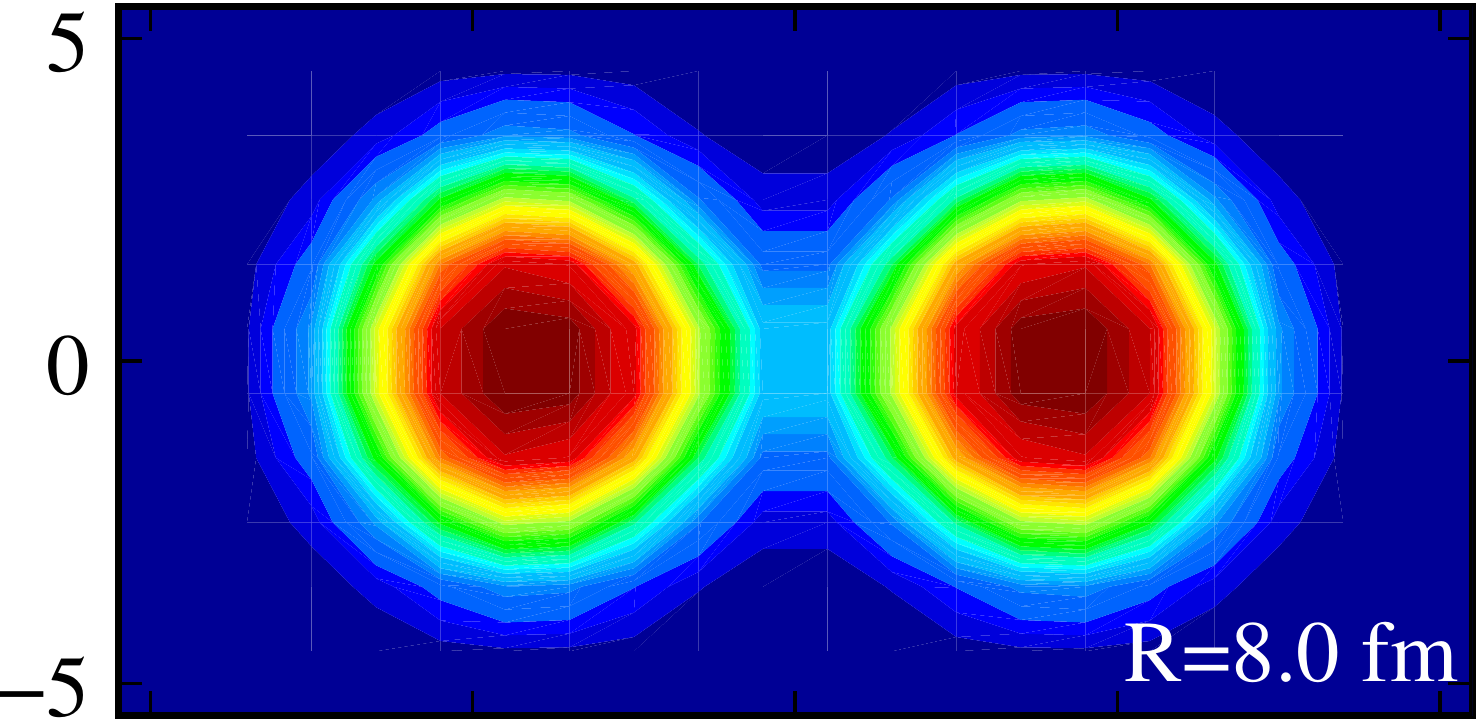}\vspace{-2pt}
\includegraphics*[width=6.6cm]{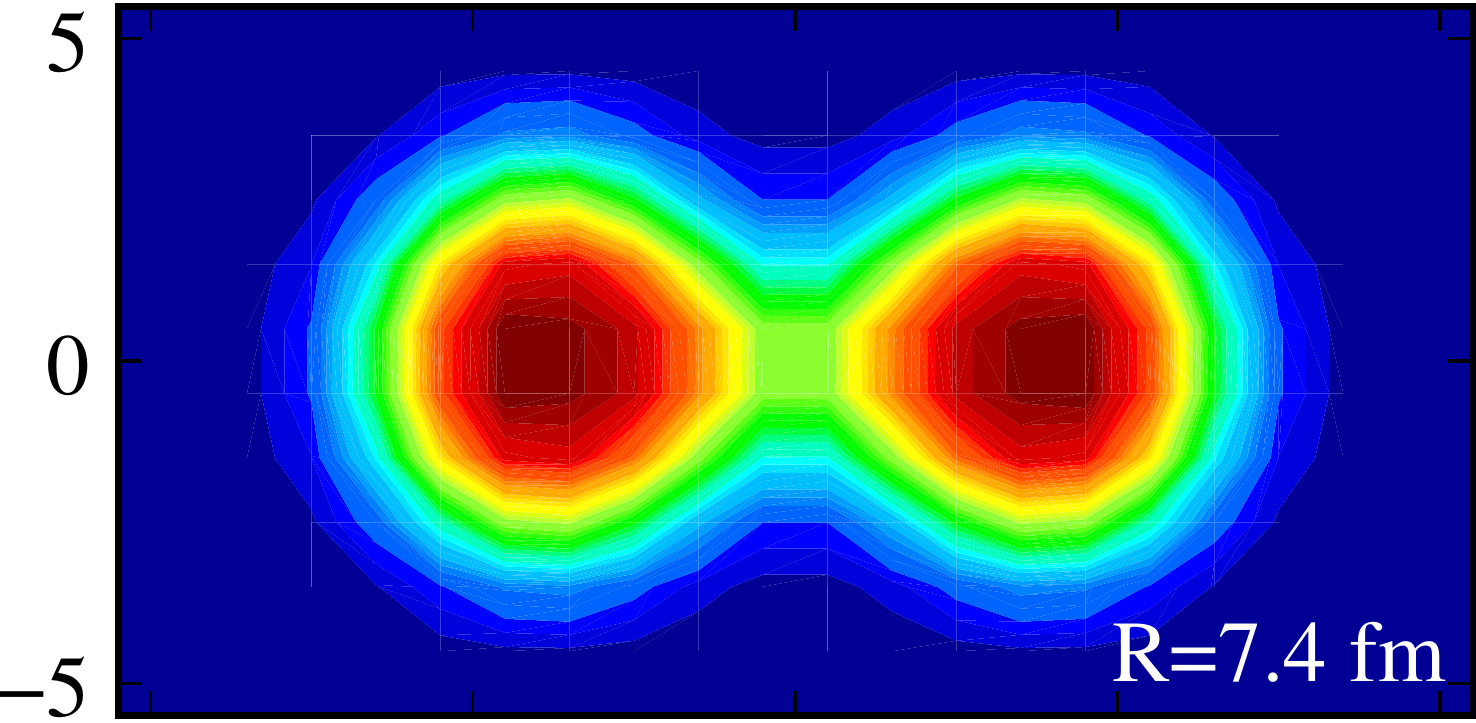}\vspace{-2pt}
\includegraphics*[width=6.66cm]{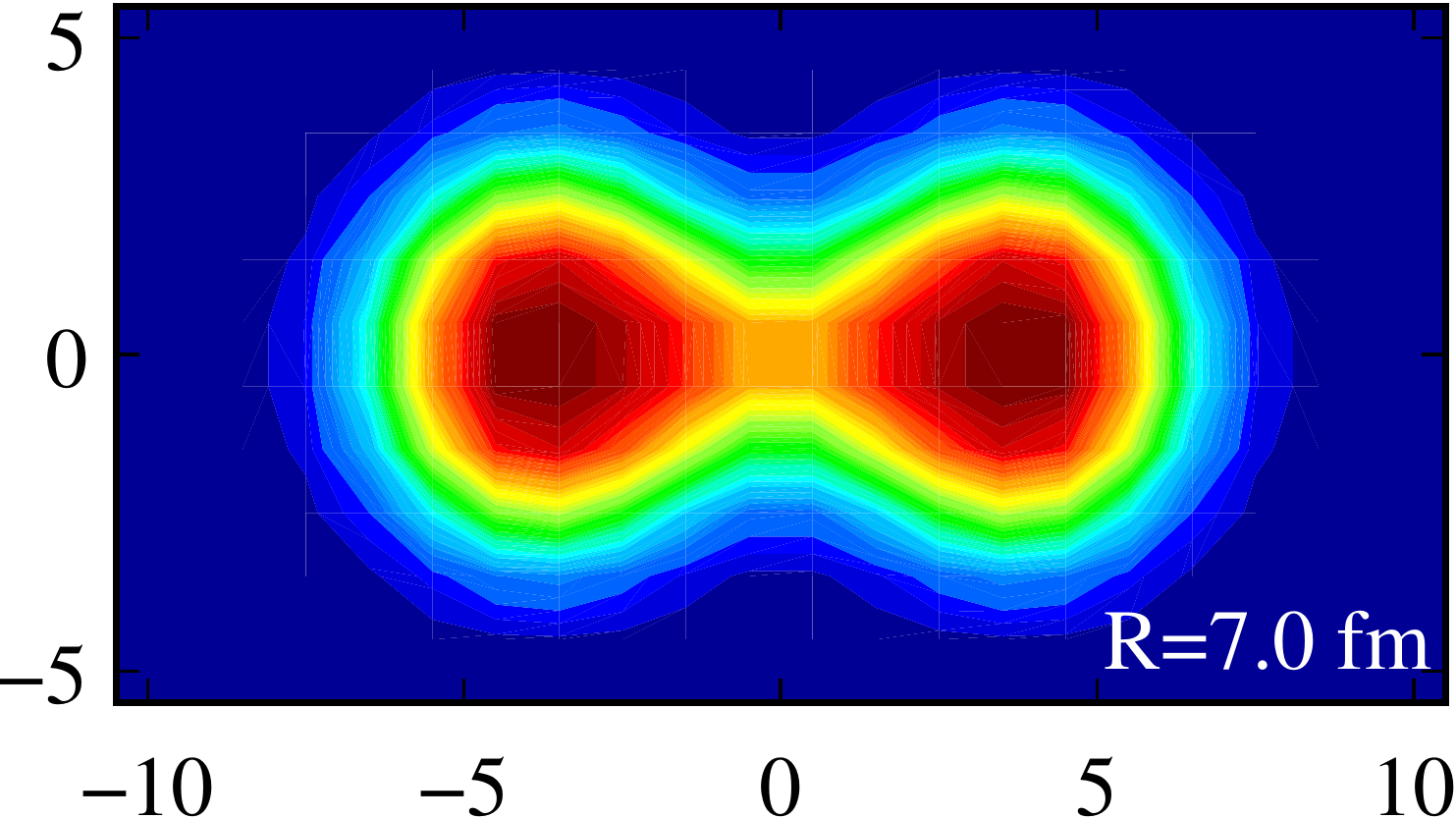}\hspace{-1pt}\vspace{-2pt}
\caption{(Color online) Snapshots of the nuclear density contours in the $x-z$ plane
during the TDHF time-evolution of the $^{16}$O+$^{16}$O system at a collision
energy $E_\mathrm{TDHF}=12$~MeV are shown for four different ion-ion separation
values of $R$. The top panel corresponds to the peak of the potential barrier at $10.0$~MeV, the subsequent
panels show the nuclear density at the inner turning point of the potential barrier
corresponding to the c.m. energies, $9.4$, $6.5$, and $3.7$~MeV, respectively.}
\label{fig1}
\end{figure}

The fusion barrier penetrabilities $T_L(E_{\mathrm{c.m.}})$
are obtained by numerical integration of the two-body Schr\"odinger equation
\begin{equation}
\left[ \frac{-\hbar^2}{2\mu}\frac{d^2}{dR^2}+\frac{L(L+1)\hbar^2}{2\mu R^2}+V(R)-E\right]\psi=0\;,
\label{eq:xfus}
\end{equation}
using the {\it incoming wave boundary condition} (IWBC) method~\cite{Raw64}.
IWBC assumes that once the minimum of the potential is reached fusion will
occur. In practice, the Schr\"odinger equation is integrated from the potential
minimum, $R_\mathrm{min}$, where only an incoming wave is assumed, to a large asymptotic distance,
where it is matched to incoming and outgoing Coulomb wavefunctions. The barrier
penetration factor, $T_L(E_{\mathrm{c.m.}})$ is the ratio of the
incoming flux at $R_\mathrm{min}$ to the incoming Coulomb flux at large distance.
Here, we implement the IWBC method exactly as it is
formulated for the coupled-channel code CCFULL described in Ref.~\cite{HR99}.
This gives us a consistent way for calculating cross-sections at above and below
the barrier via
\begin{equation}
\sigma_f(E_{\mathrm{c.m.}}) = \frac{\pi}{k^2} \sum_{L=0}^{\infty} (2L+1) T_L(E_{\mathrm{c.m.}})\;.
\label{eq:sigfus}
\end{equation}

\begin{figure}[!htb]
\includegraphics*[width=6.6cm]{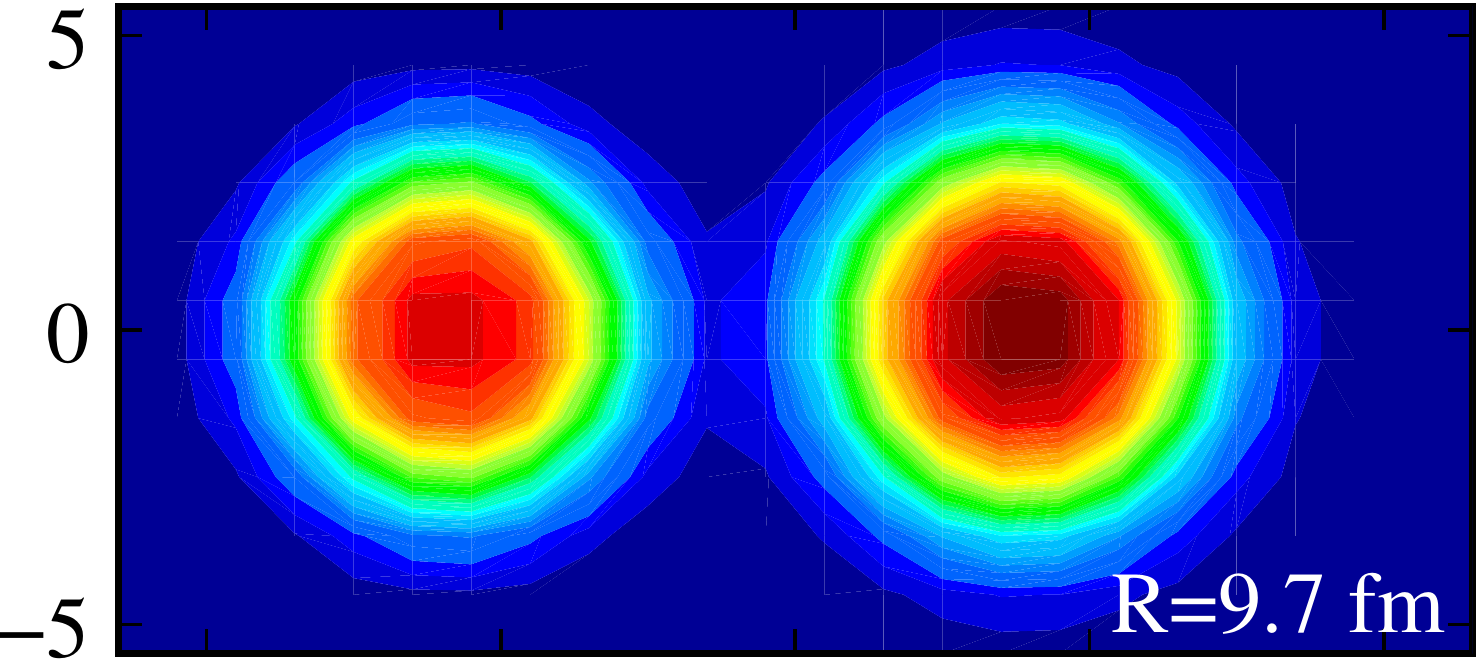}\vspace{-2pt}
\includegraphics*[width=6.6cm]{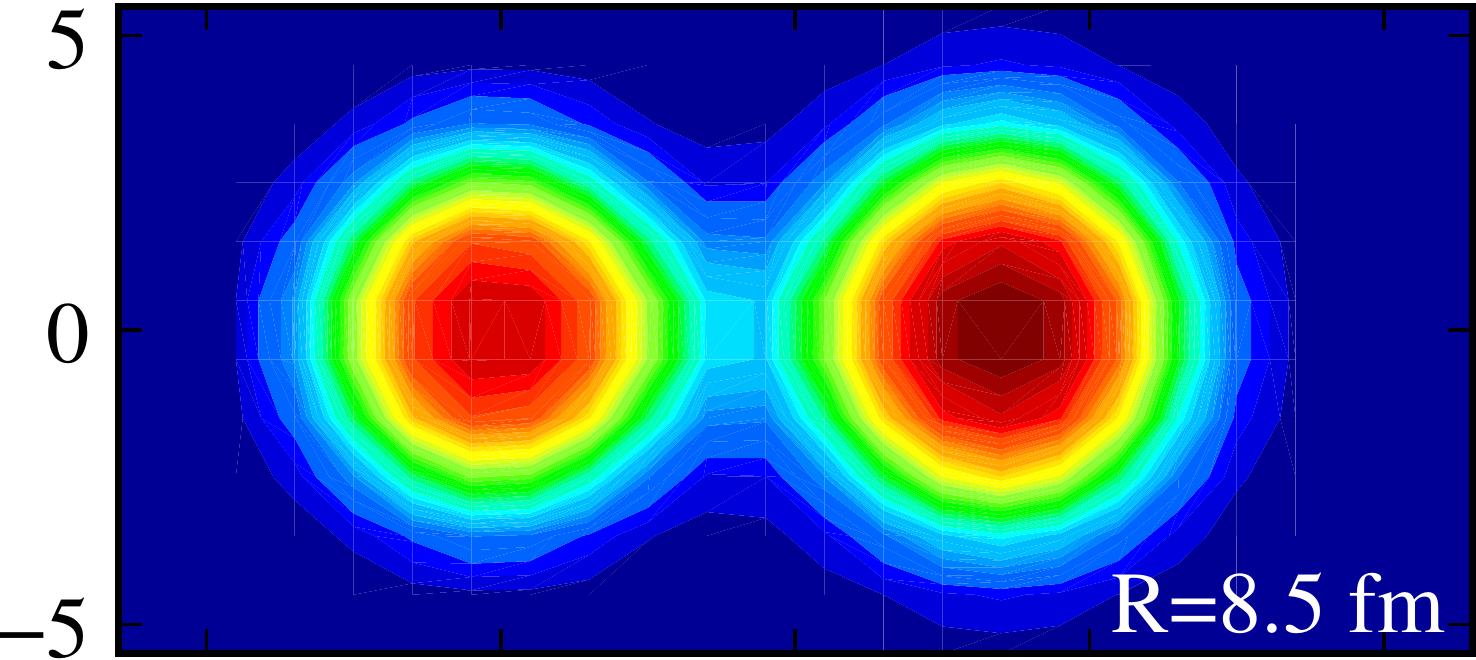}\vspace{-2pt}
\includegraphics*[width=6.6cm]{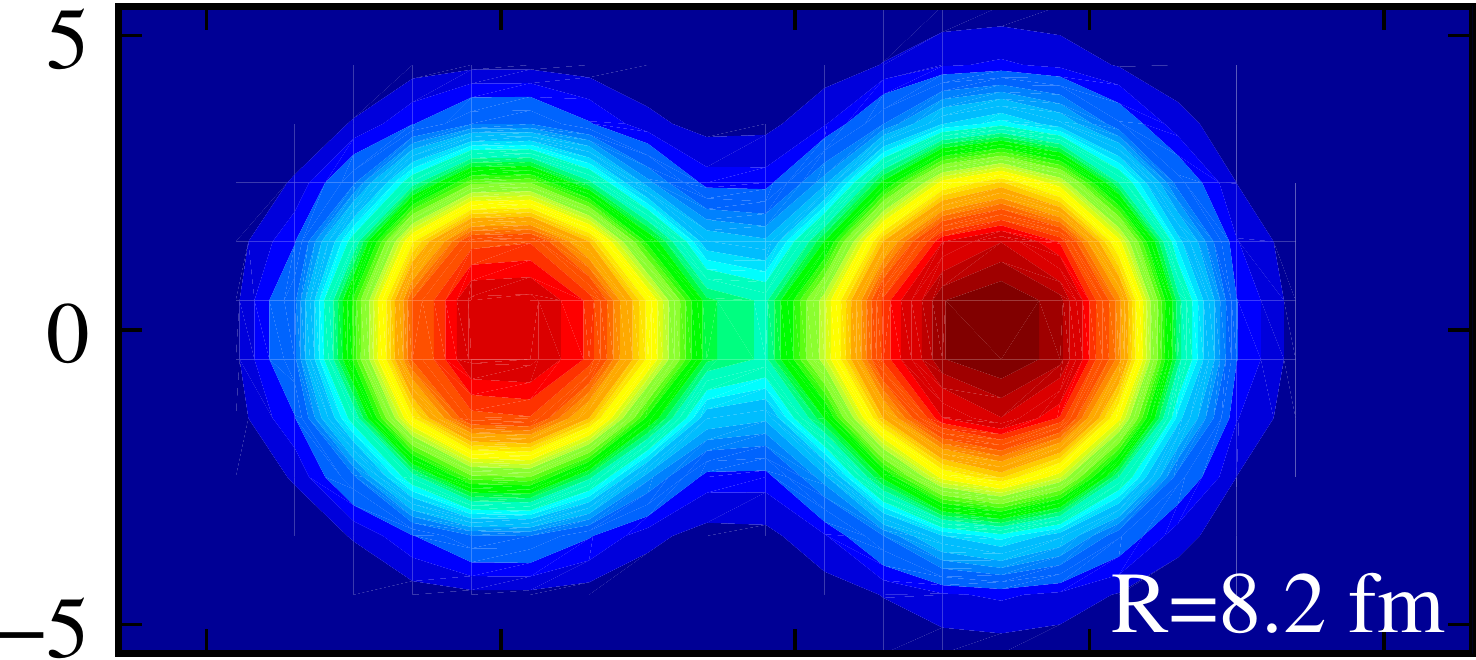}\vspace{-2pt}
\includegraphics*[width=6.6cm]{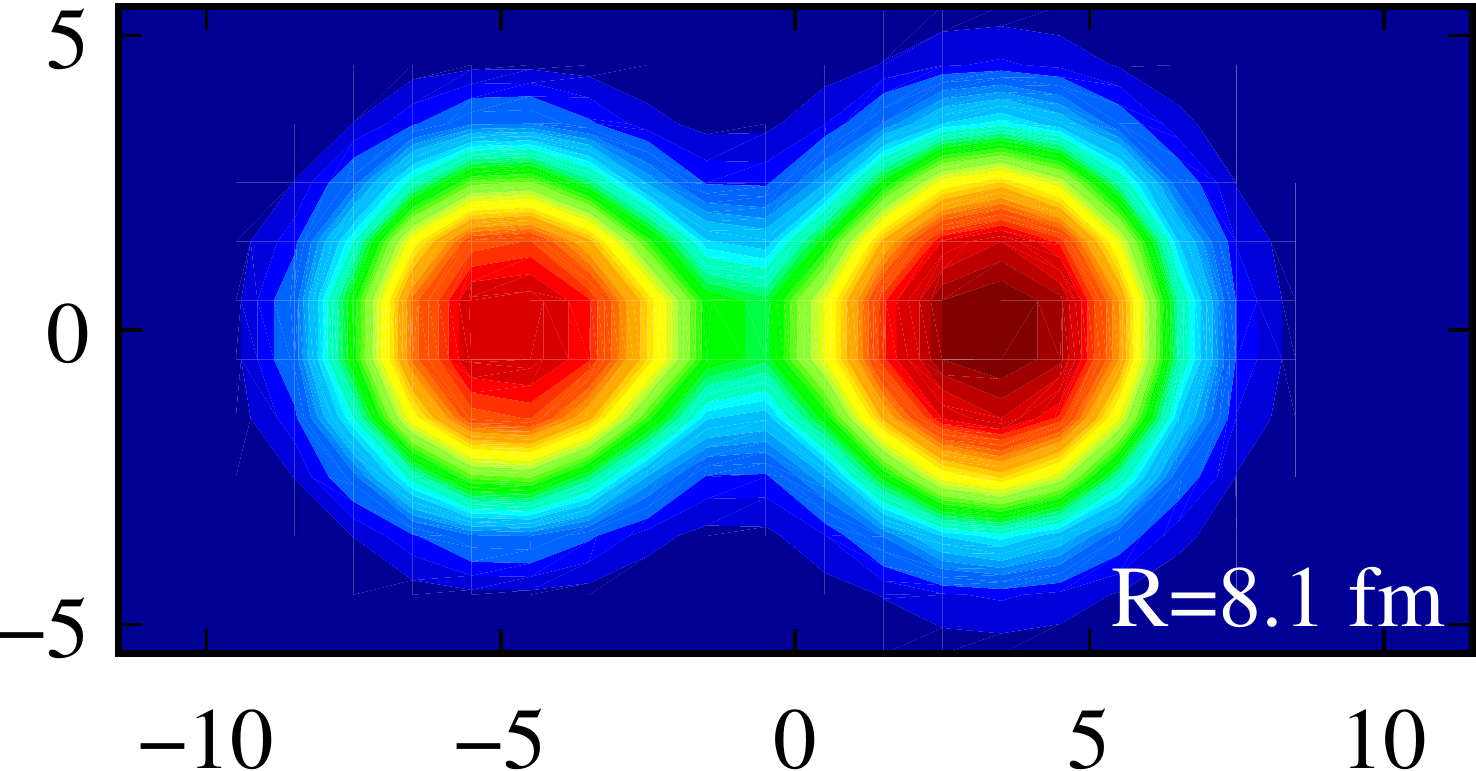}\hspace{-0pt}\vspace{-2pt}
\caption{(Color online) Snapshots of the nuclear density contours in the $x-z$ plane
during the TDHF time-evolution of the $^{16}$O+$^{24}$O system
at a collision energy $E_\mathrm{TDHF}=12$~MeV are shown for four different ion-ion separation
values of $R$. The top panel corresponds to the peak of the potential barrier at $8.5$~MeV, the subsequent
panels show the nuclear density at the inner turning point of the potential barrier
corresponding to the c.m. energies, $6.0$, $4.0$, and $3.0$~MeV, respectively.}
\label{fig2}
\end{figure}

\subsection{Simple S\~{a}o Paulo barrier penetration model}
\label{subsec.saopaulo}

In this section we describe the simple S\~{a}o Paulo barrier penetration model to calculate fusion cross-sections. This
starts with the double folding potential $V_F(R)$~\cite{sp},
\begin{equation}
V_F(R)=\int d^3r_1d^3r_2 \rho_1(r_1)\rho_2(r_2) V_0 \delta(\mathbf{r}_1- \mathbf{r}_2 - \mathbf{R})\, .
\label{vf}
\end{equation}
Here $\rho_1$ and $\rho_2$ are the ground-state densities of the two nuclei and $V_0=-456$ MeV-fm$^{3}$.
From $V_F$ a nonlocal potential $V(R,E)$ is constructed, $V(R,E)=V_F(R)e^{-4v^2/c^2}$, where $v$ is the local relative
velocity~\cite{sp} and $c$ is the speed of light.  Finally, tunneling through the Coulomb plus $V(R,E)$ potentials is
calculated in a WKB approximation to infer the fusion cross section~\cite{sp, pycno}.

\section{Results}
\label{sec.results}
In this section we present results for effective potentials and fusion cross-sections using the
two approaches outlined in the preceding section.

Time-dependent Hartree-Fock calculations were done in 3-D geometry and using the full Skyrme interaction
including all of the time-odd terms in the mean-field Hamiltonian~\cite{UO06}.
The Skyrme parametrization used was SLy4~\cite{CB98}.  In addition to providing a good description of nuclei, this interaction has been used to produce an equation of state for neutron stars \cite{Douchin}.
For the
reactions studied here, the lattice spans $44$~fm along the collision axis and $22$~fm in
the other two directions. Derivative operators on
the lattice are represented by the Basis-Spline collocation method. One of the major
advantages of this method is that we
may use a relatively large grid spacing of $1.0$~fm and nevertheless achieve high numerical
accuracy.
The initial separation of the two nuclei was $18$~fm for central collisions and the TDHF collision
energy was $E_\mathrm{TDHF}=12$~MeV.
The time-propagation
is carried out using a Taylor series expansion (up to orders $10-12$) of the unitary mean-field propagator,
with a time-step $\Delta t = 0.4$~fm/c.
The accuracy of the density constraint calculations is
commensurate with the accuracy of the static calculations.

Calculations were done on a $16$-processor Linux workstation using our OpenMP TDHF code with
almost a $100$\% parallel efficiency.
We have performed density constraint calculations every $10-20$ time steps.
For the light systems considered here, one full calculation takes about $12$ hours.

In Fig.~\ref{fig1} we plot snapshots of the nuclear density contours in the $x-z$ plane
during the TDHF time-evolution of the $^{16}$O+$^{16}$O system at four different ion-ion separation
values of $R$. The top panel corresponds to the peak of the potential barrier at $10.0$~MeV, the subsequent
panels show the nuclear density at the inner turning point of the potential barrier
corresponding to c.m. energies of $9.4$, $6.5$, and $3.7$~MeV, respectively.
As one can observe from these densities for low c.m. energies the nuclei have a substantial overlap
and develop a {\it neck} which is very different from two overlapping spherical densities as in the
frozen-density approach. Figure~\ref{fig2} shows the density contours for the asymmetric
$^{16}$O+$^{24}$O system. Again, the top panel represents the density contours corresponding to the peak of the
corresponding potential barrier at $8.5$~MeV, the subsequent panels show the same quantity
at barrier energies $6.0$, $4.0$, and $3.0$~MeV, respectively.
The difference between the two figures is interesting. For the neutron-rich system
the barrier peak is at a larger $R$ value since the nuclei come into contact sooner
due to the extended neutron skin of the $^{24}$O nucleus.

\begin{figure}[!htb]
\begin{center}
\includegraphics*[width=8.6cm] {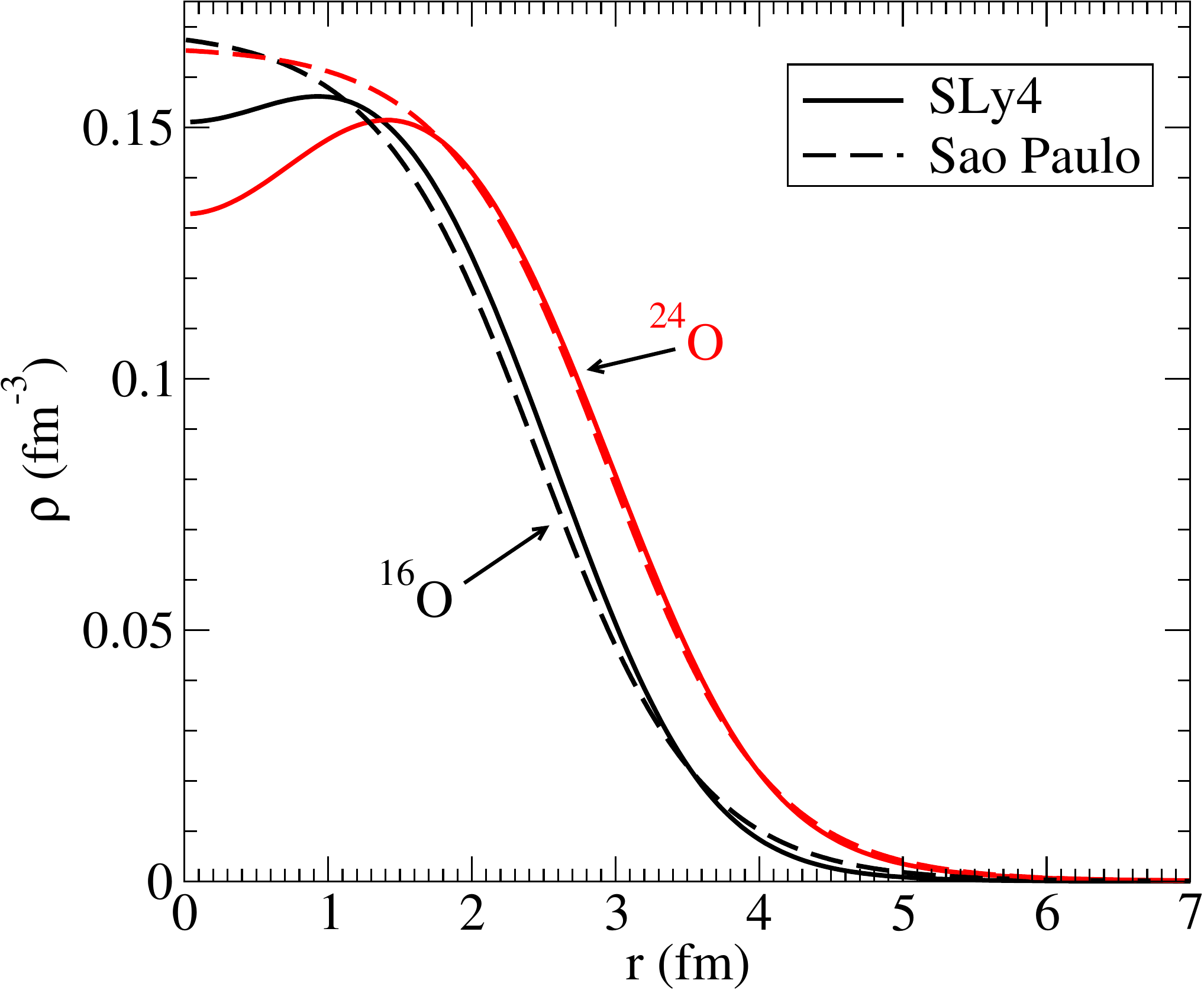}
\caption{(Color online) Baryon density of oxygen isotopes $^{16}$O (black) and $^{24}$O (red) versus radial
distance $r$.  The
solid lines are HF results for the SLy4 interaction while the dashed lines are simple parameterized Woods-Saxon densities
used in the S\~{a}o Paulo barrier penetration model, see Sec.~\ref{subsec.saopaulo}.}
\label{fig3}
\end{center}
\end{figure}

For simplicity the S\~{a}o Paulo model often assumes Woods-Saxon static nuclear densities with radius parameter $R=1.31A^{1/3}-0.84$~fm and
diffuseness $a=0.58$~fm~\cite{sp,pycno}.   However there are also S\~{a}o Paulo model results using mean field densities \cite{M.Beard}.  In Fig.~\ref{fig3} we compare these Woods-Saxon ground state densities with those
calculated with a spherical Hartree-Fock program using the Skyrme SLy4 interaction.
As one can see the surface properties calculated with the two-approaches agree reasonably well, whereas the
central dip in the density, characteristic to some light nuclei, is not reproduced by the Woods-Saxon shape.

\begin{figure}[!htb]
\begin{center}
\includegraphics*[width=8.6cm] {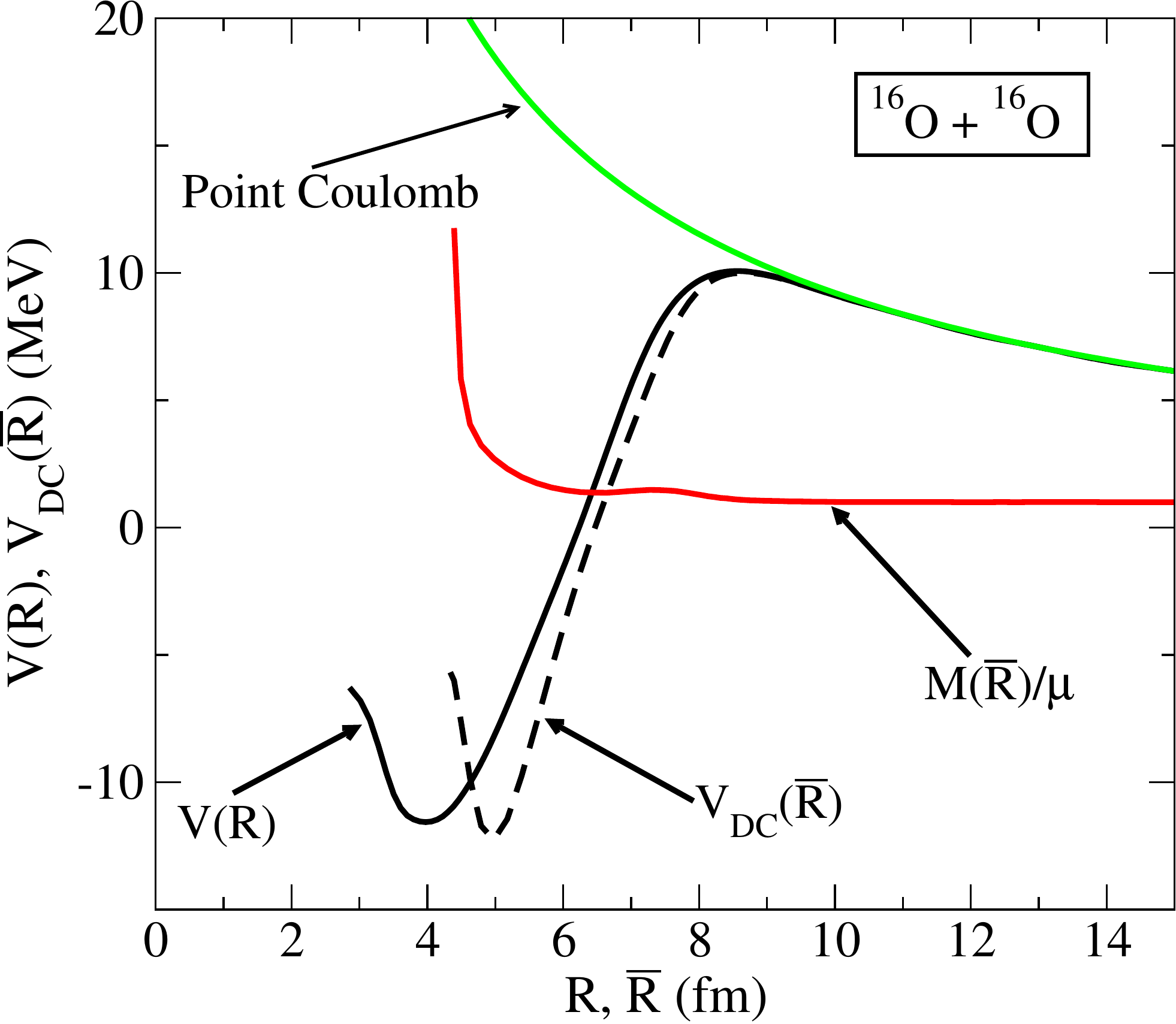}
\caption{(Color online) DC-TDHF potential barriers for the  $^{16}$O$+$ $^{16}$O system.
The dashed line shows the potential $V_{DC}(\bar{R})$ calculated directly from
TDHF with density constraints. The coordinate-dependent mass $M(\bar{R})/\mu$ is also
shown on the same plot. The point transformation Eq.~(\ref{eq:mrbar}) results in the
potential $V(R)$. Also shown is the point Coulomb potential.}
\label{fig4}
\end{center}
\end{figure}

In Fig.~\ref{fig4} we have plotted various quantities discussed in Sec.~\ref{subsec.TDHF}
for the $^{16}$O$+$ $^{16}$O  system. We observe how the coordinate-dependent mass modifies
the inner part of the barrier when the point transformation is performed.
The dashed line shows the potential $V_{DC}(\bar{R})$ calculated directly from
TDHF with density constraint. The coordinate-dependent mass $M(\bar{R})/\mu$ is also
shown on the same plot. As we can see this ratio equals one when the two nuclei are
far apart and starts to deviate as the nuclei slow down and finally peaks when
$\dot{\bar{R}}$ approaches zero.
The point transformation Eq.~(\ref{eq:mrbar}) results in the
potential $V(R)$.
We stress that the transformation given in Eq.~(\ref{eq:mrbar}) is carried out
for numerical convenience: The original Schr\"{o}dinger equation
which involves the coordinate-dependent mass $M(\bar{R})$ and the
potential $V_{DC}(\bar{R})$ is mathematically equivalent to the
transformed Eq.~(\ref{eq:xfus}) which contains
the constant reduced mass $\mu$ and the transformed potential
$V(R)$. However, from a numerical point of view, Eq.~(\ref{eq:xfus}) is
easier to solve because of the simpler structure of the kinetic energy operator.

\begin{figure}[!htb]
\begin{center}
\includegraphics*[width=8.6cm] {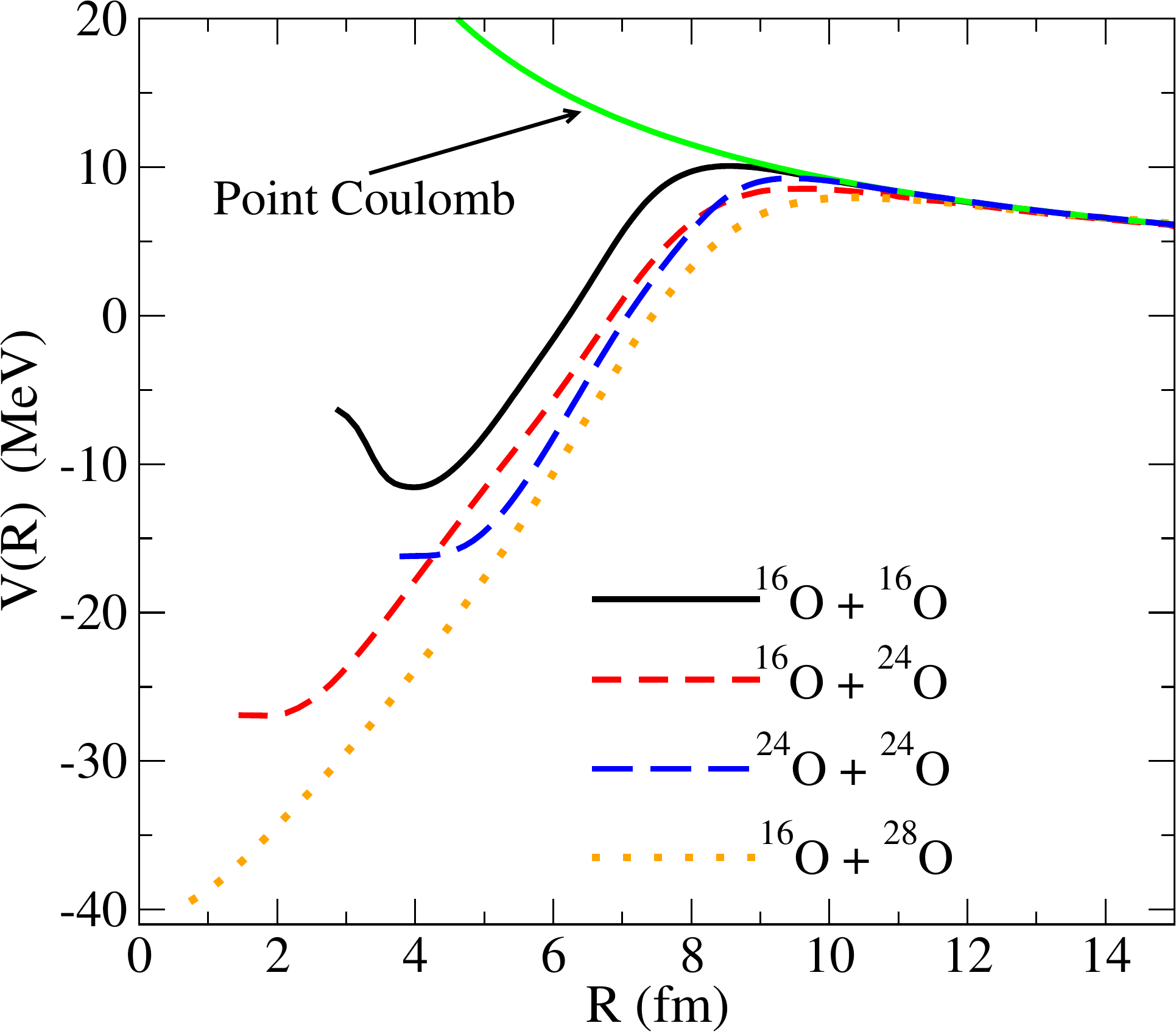}
\caption{(Color online) Potential barriers for the O$+$O system obtained from density constrained
TDHF calculations, $^{16}$O$+$ $^{16}$O (black solid curve), $^{16}$O$+$ $^{24}$O
(red short-dash curve), $^{24}$O$+$ $^{24}$O (blue long-dash curve), and $^{16}$O$+$ $^{28}$O (orange dotted curve).
Also shown is the point Coulomb potential.}
\label{fig5}
\end{center}
\end{figure}

In Fig.~\ref{fig5} we plot the ion-ion potentials obtained using the density-constrained TDHF
approach for various oxygen isotopes. These potentials include all the dynamical effects
as well as the coordinate dependent mass. We see that the $^{16}$O$+$ $^{16}$O
potential (black solid curve) has the highest barrier of about $10$~MeV. As the nuclear mass increases the nuclei
come into contact earlier, which corresponds to a larger $R$ value, and deviate from the
point Coulomb potential (green solid curve), consequently having a lower barrier height.
The barrier heights in decreasing order are $10.00$, $9.24$, $8.54$, and $7.95$~MeV.
The reason for potentials extending to smaller $R$ values in the inner part of the barrier
is due to the fact that the collision energy is effectively higher for lower barriers
thus the system can reach a more compact shape before fusion.
With the SLy4 interaction the $^{28}$O nucleus is barely bound, while the experimental drip
line is at $^{24}$O, so we have included this nucleus as well.

\begin{figure}[!htb]
\begin{center}
\includegraphics*[width=8.6cm] {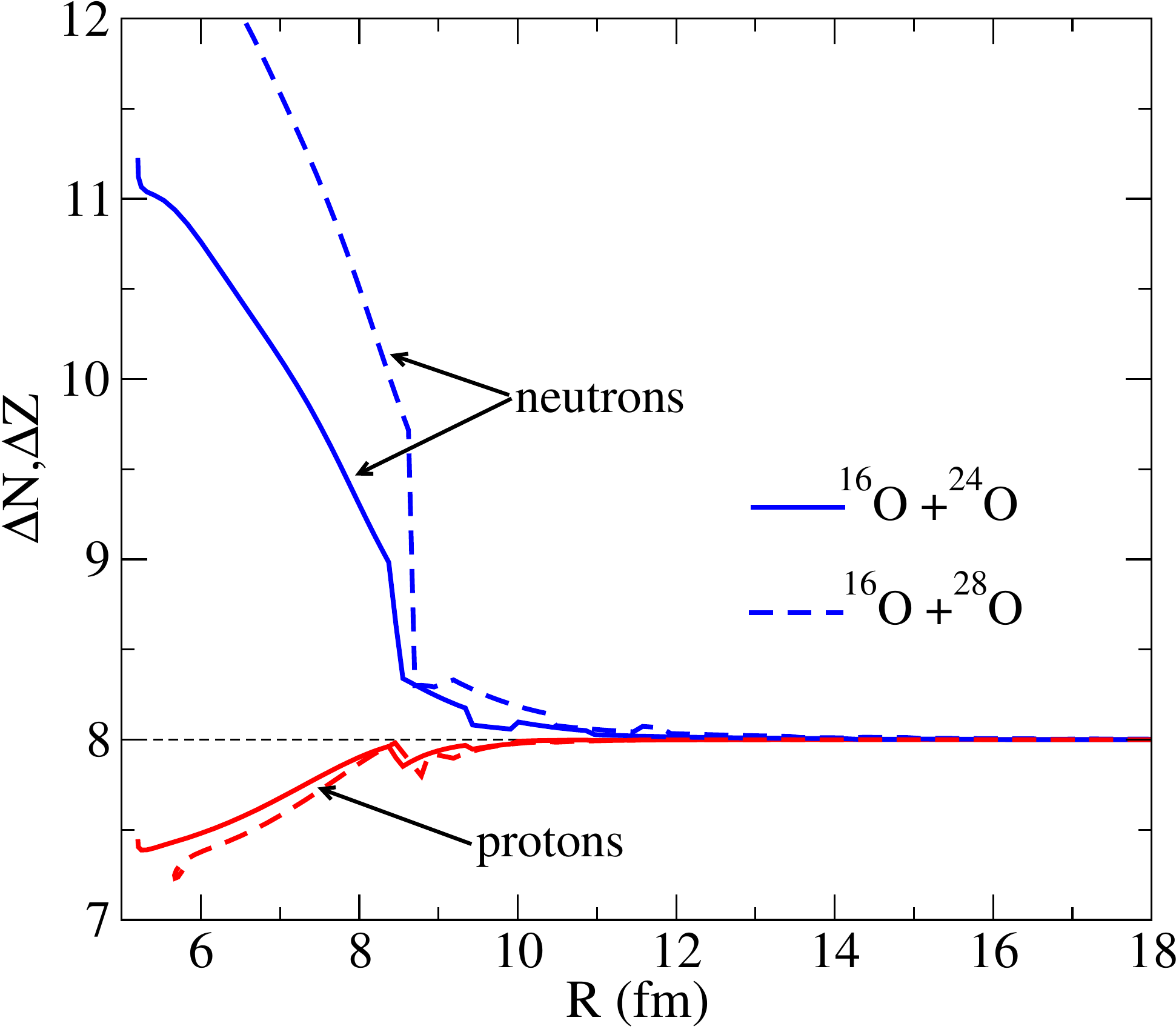}
\caption{(Color online) Neutron transfer (blue) and proton transfer (red) to $^{16}$O from
the neutron-rich reaction partner versus separation $R$ during the fusion of
$^{16}$O on $^{24}$O (solid lines) and on $^{28}$O (dashed lines).}
\label{fig6}
\end{center}
\end{figure}
Excitations are believed to have a significant impact on the outcome of the
fusion reactions. The excitations can range from the entrance channel
quantal excitations of the projectile and target, as in the coupled-channel
approach, to collective excitations of pre-equilibrium system, to compound
nucleus excitations. These can be further influenced by particle transfer,
pre-equilibrium emissions, and evaporation, among others. Theoretically such
effects are commonly introduced by hand into various reaction models.
However, the influence of excitations on nuclear reaction dynamics remains
to be a difficult and an open problem as it combines both nuclear structure
and dynamics under nonequilibrium conditions.
In Fig.~\ref{fig6} we plot the average number of neutrons and protons
transferred to the $^{16}$O nucleus during the early stages of the TDHF collisions
for the $^{16}$O+$^{24}$O (solid curves)
and the $^{16}$O+$^{28}$O (dashed curves) systems.
Neutrons are denoted by blue curves and
protons by red curves.
These transfers are calculated using the standard method in TDHF, which is to
take a cut at the point of minimum density in the overlap region between the
two nuclei and integrate the densities in the left and right halves of the
collision box.
A number of interesting things can be observed
from this plot; first is the fact that most of the transfer seems to
start after we pass the potential barrier peaks. This indicates that
particle transfer primarily modifies the inner part of the barriers
and not so much the barrier heights. The other observation is that
on average about one neutron is transferred from the $^{24}$O to $^{16}$O in the region of $R$ values
relevant for fusion cross-sections, whereas two neutrons are transferred from $^{28}$O to $^{16}$O
in the same region. We also note that these transfers occur rather rapidly.
\begin{figure}[!htb]
\begin{center}
\includegraphics*[width=8.6cm] {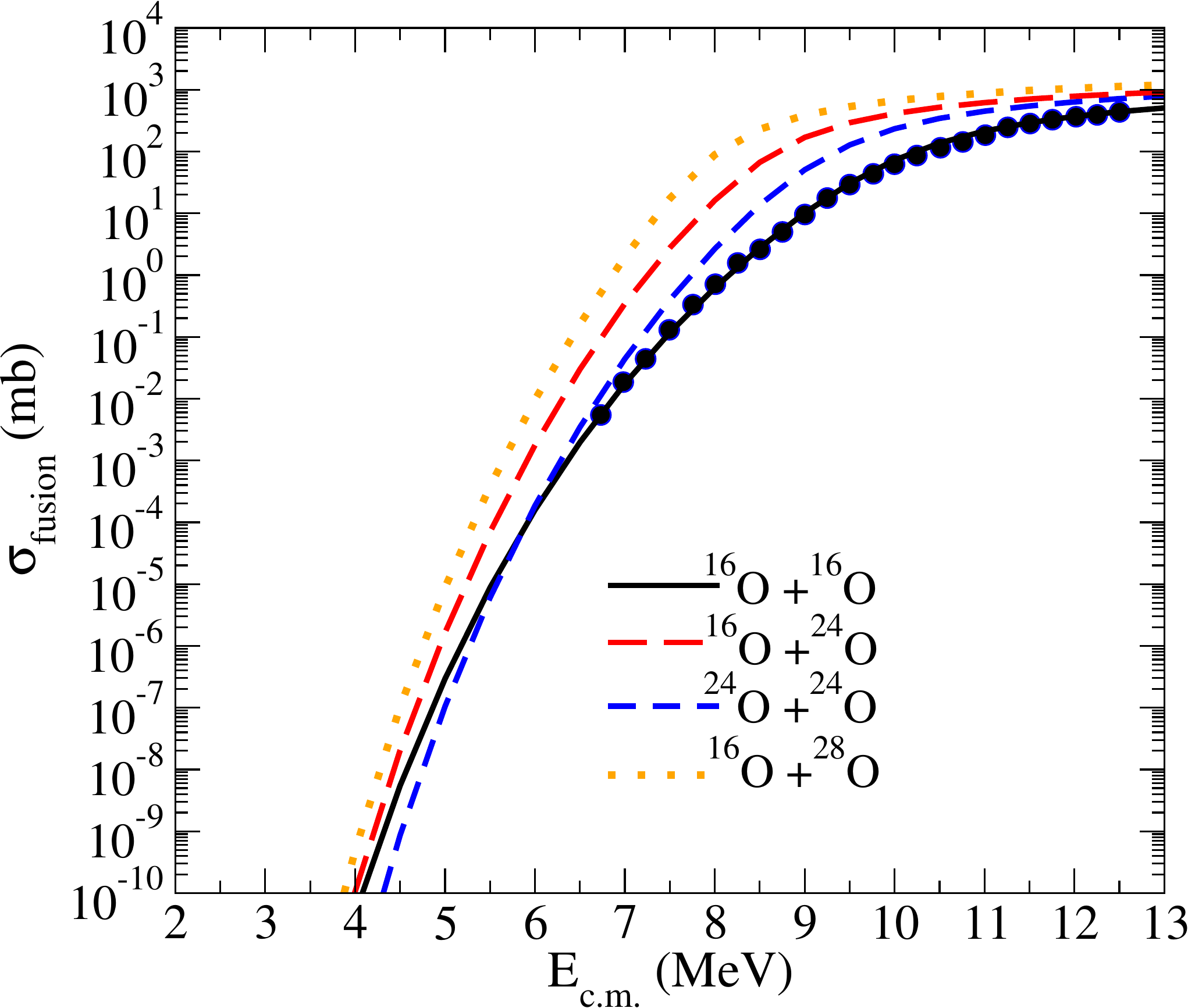}
\caption{(Color online) Cross-section versus center-of-mass energy for fusion of oxygen isotopes.
Experimental results (filled circles) are from Ref.~\cite{o16o16expdata_2}}
\label{fig7}
\end{center}
\end{figure}

We have obtained the fusion cross-sections for the O$+$O systems by numerical integration of Eq.~(\ref{eq:xfus})
to calculate the transmission probability and Eq.~(\ref{eq:sigfus}) for the fusion cross sections.
The resulting cross-sections are shown in Fig.~\ref{fig7}.
We observe that for the $^{16}$O+$^{16}$O system barriers we obtain a very good description of the
low-energy experimental fusion cross-sections. The higher energy part of the fusion cross-sections are
primarily determined by the barrier properties in the vicinity of the barrier
peak. On the other hand sub-barrier cross-sections are influenced by what
happens in the inner part of the barrier and here the dynamics and
consequently the coordinate dependent mass becomes very important.
Also shown in Fig.~\ref{fig7} are the fusion cross-sections for $^{16}$O+$^{24}$O, $^{16}$O+$^{28}$O,
and the $^{24}$O+$^{24}$O systems. The behavior of these cross-sections can be deduced from the potential
barriers of Fig.~\ref{fig5}. The highest fusion cross-sections belong to the collision of $^{16}$O with
the most neutron rich isotope $^{28}$O. The interesting observation is that the collisions involving
$^{16}$O and a neutron-rich isotope seem to have a larger fusion yield than the $^{24}$O+$^{24}$O system.
\begin{figure}[!htb]
\begin{center}
\includegraphics*[width=8.6cm] {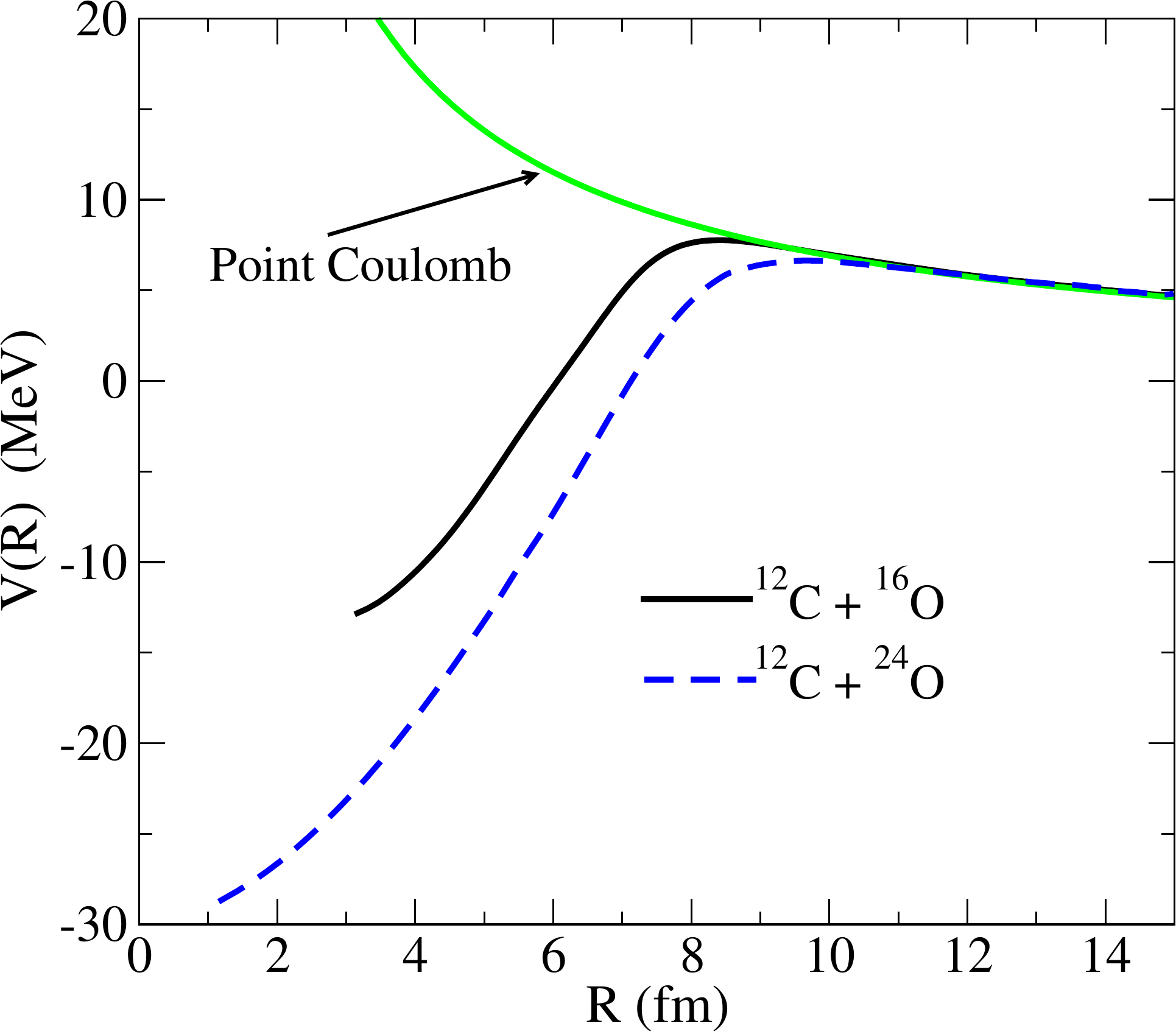}
\caption{(Color online)  Potential barriers for the C$+$O system obtained from density constrained
TDHF calculations, $^{12}$C$+$ $^{16}$O (black solid curve) and $^{12}$C$+$ $^{24}$O
(blue dashed curve). Also shown is the point Coulomb potential (solid green curve).}
\label{fig8}
\end{center}
\end{figure}

In Fig.~\ref{fig8} we show the DC-TDHF potential barriers for the C$+$O system.
The higher barrier corresponds to the $^{12}$C$+$ $^{16}$O system and has a peak
energy of
$7.77$~MeV. The barrier for the $^{12}$C$+$ $^{24}$O system occurs at a
slightly larger $R$ value with a barrier peak of $6.64$~MeV.
\begin{figure}[!htb]
\begin{center}
\includegraphics*[width=8.6cm] {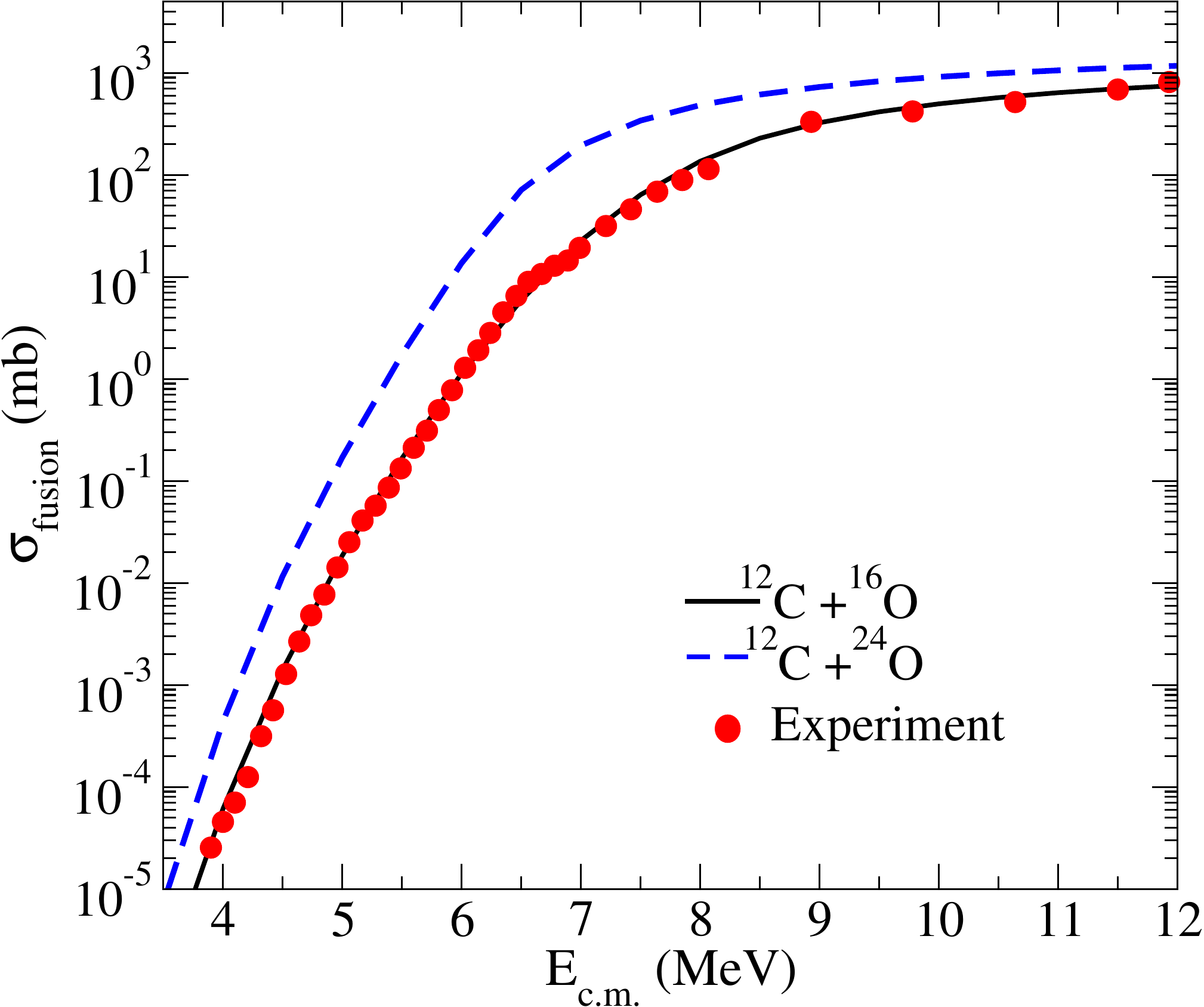}
\caption{(Color online) Cross-section versus center-of-mass energy for fusion of $^{12}$C with oxygen
isotopes.  Experimental data (circles) are from Ref.~\cite{c12o16expdata_3}.}
\label{fig9}
\end{center}
\end{figure}
Figure~\ref{fig9} shows the corresponding cross-sections for the two reactions.
Also shown are the experimental data from Ref.~\cite{c12o16expdata_3}. The
DC-TDHF potential reproduces the experimental cross-sections quite well for the
$^{12}$C$+$ $^{16}$O system.  Again the cross section for the neutron rich $^{12}$C+$^{24}$O is seen to be larger than
that for $^{12}$C +$^{16}$O.

Some of the strong energy dependence of the fusion cross section $\sigma(E)$ can be taken into account with the
Astrophysical $S$ factor
\begin{equation}
S(E) = \sigma(E) E\, {\rm Exp}[2\pi\eta],
\end{equation}
at center-of-mass energy $E$ and Sommerfeld parameter $\eta=Z_1Z_2e^2/(\hbar v)$.  Finally, the relative velocity of
the nuclei is $v=\sqrt{2E/\mu}$ for a system of reduced mass $\mu$.

\begin{figure}[!htb]
\begin{center}
\includegraphics*[width=8.6cm] {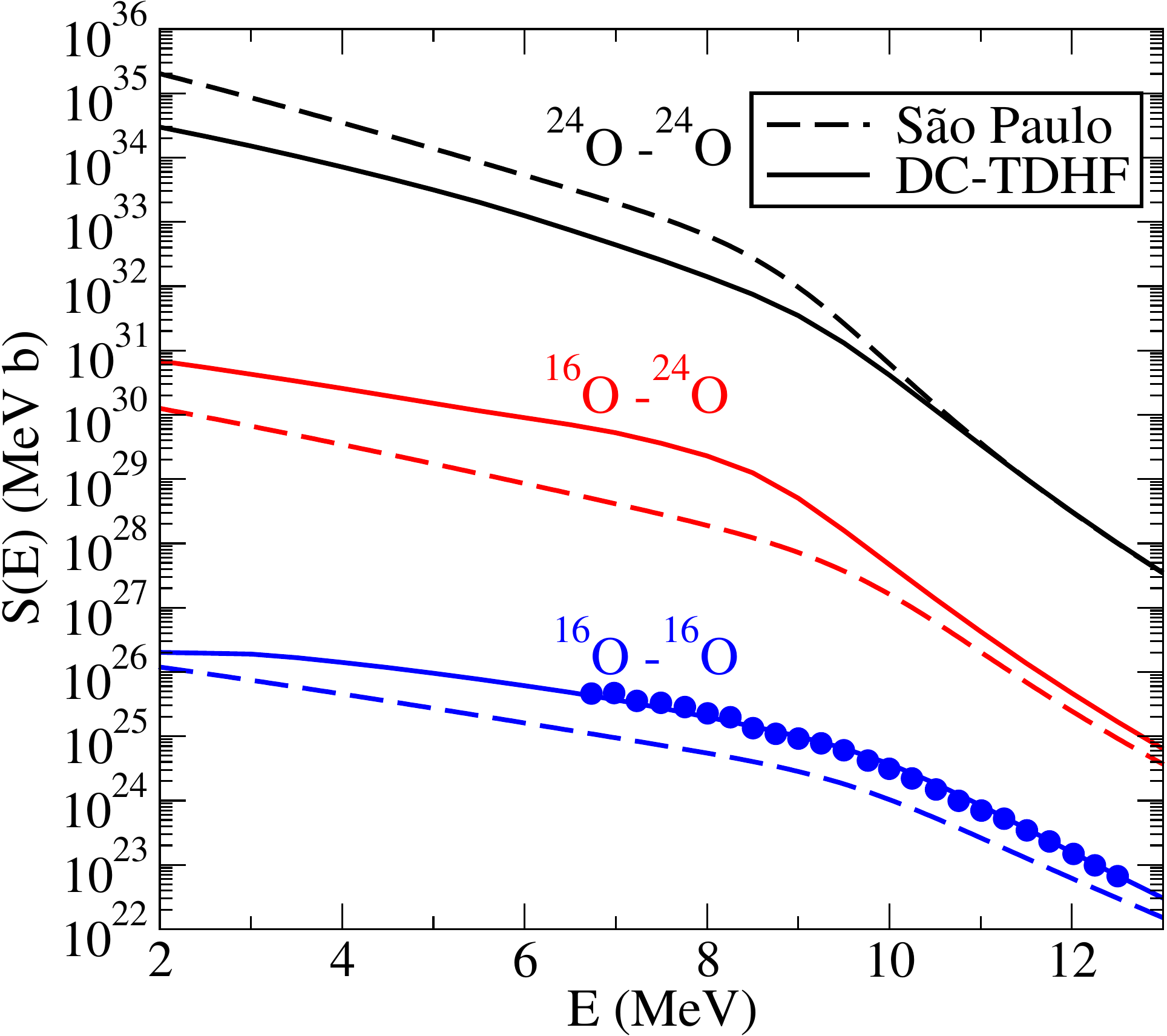}
\caption{(Color online) Astrophysical $S$ factor versus center of mass energy for fusion of oxygen isotopes.  Solid
lines show TDHF results while dashed lines are for the S\~{a}o Paulo barrier penetration model.  Experimental results
(circles) are from ref.~\cite{o16o16expdata_2}.}
\label{fig10}
\end{center}
\end{figure}

\begin{figure}[!htb]
\begin{center}
\includegraphics*[width=8.6cm] {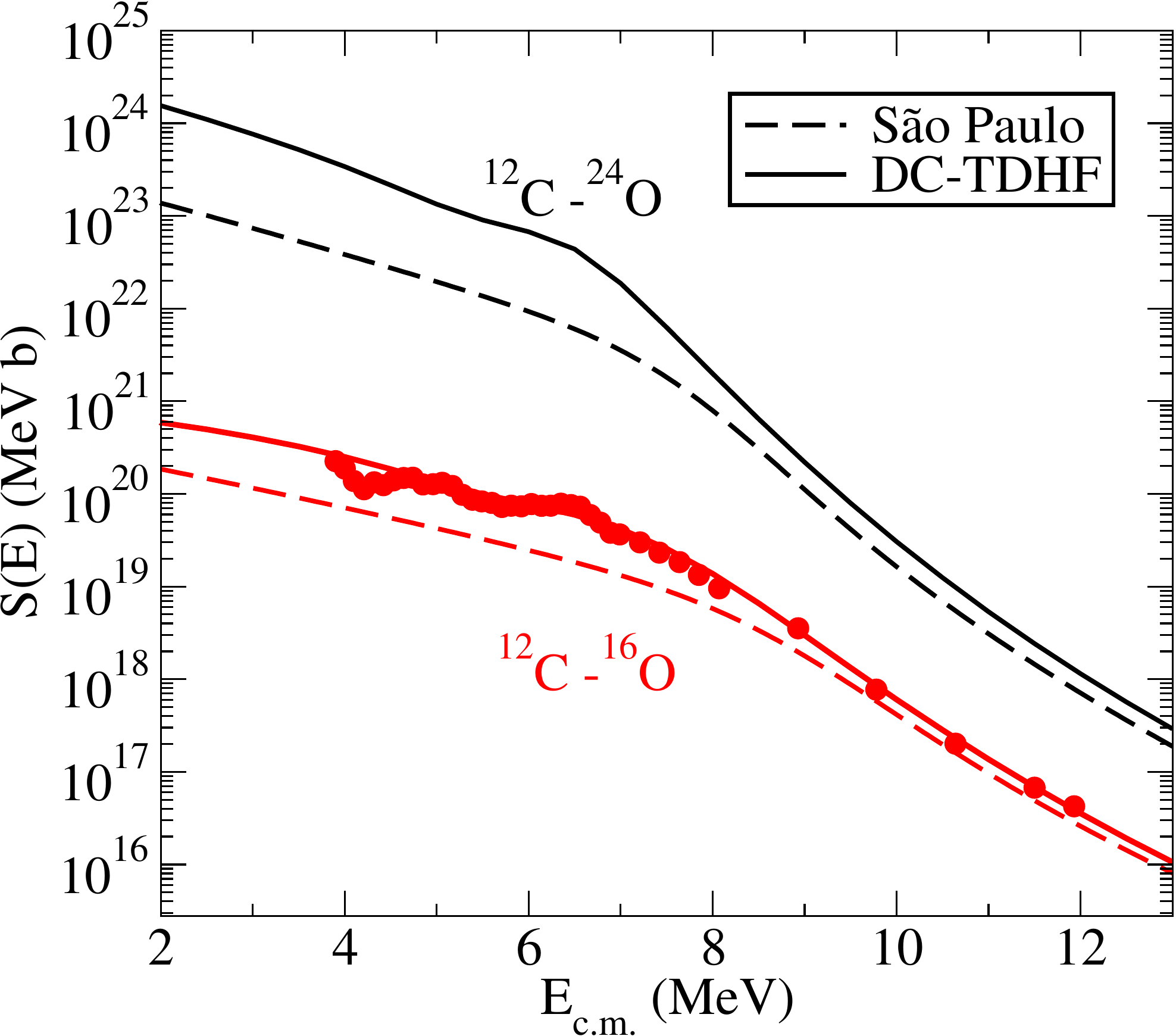}
\caption{(Color online) Astrophysical $S$ factor versus center of mass energy for fusion of $^{12}$C with oxygen
isotopes.  Solid lines show TDHF results while dashed lines are for the S\~{a}o Paulo barrier penetration model.
Experimental results (circles) are from ref.~\cite{c12o16expdata_3}.}
\label{fig11}
\end{center}
\end{figure}

The $S$ factor for our TDHF calculations is shown in Figs.~\ref{fig10} and \ref{fig11}. Also shown are $S$ factors from
accurate nine parameter fits to S\~{a}o Paulo model results~\cite{M.Beard} (dashed lines). In Fig.~\ref{fig10} the TDHF $S$
factor for $^{16}$O+$^{16}$O agrees well with data.  Presumably a small change in the S\~{a}o Paulo potential will allow the
S\~{a}o Paulo results to also agree well with data.  However, there are interesting differences for more neutron rich
systems.  We note that the S\~{a}o Paulo $S$ factor for the symmetric system $^{24}$O+$^{24}$O is larger than the TDHF result
while the S\~{a}o Paul $S$ for the asymmetric system $^{16}$O+$^{24}$O is significantly smaller than the TDHF result. We
believe this pattern is due to dynamics present in the TDHF formalism that is not included in the simple S\~{a}o Paulo model.

Figure~\ref{fig11} shows a similar pattern for reactions involving $^{12}$C.  The TDHF results agree well with data for
$^{12}$C+$^{16}$O and predict a much larger $S$ factor, compared to the S\~{a}o Paulo model, for the neutron rich asymmetric
system $^{12}$C+$^{24}$O.   We discuss these differences between our TDHF results and S\~{a}o Paulo model results in the
next section.

Finally, fusion in accreting neutron stars may take place in the inner crust where there is a background gas of
neutrons.  This background gas could impact fusion cross sections.  To investigate this we are performing TDHF
simulations where the initial conditions involve both the two nuclei, appropriately boosted towards each other, and some
unbound neutrons, that are uniformly distributed in the simulation volume.  We will report these results in a later
publication.

\section{Summary and Conclusions}
\label{sec.conclusions}
Fusion of very neutron rich nuclei may be important to determine the composition and heating of the crust of accreting
neutron stars.  In this paper we calculate fusion cross sections using Time Dependent Hartree Fock simulations coupled
with density constrained Hartree Fock calculations to deduce an effective potential.  We find remarkable agreement with
experimental cross sections for the fusion of stable nuclei.  Note that our simulations use the SLy4 Skyrme force that
has been previously fit to the properties of stable nuclei. No parameters have been fit to fusion data.

We compare our results to the simple S\~{a}o Paulo barrier penetration model. This model calculates an effective potential
by folding over static densities for the projectile and target.  Within, very roughly, an order of magnitude, our
results agree with the S\~{a}o Paulo model. This provides an error estimate for Astrophysical applications of the S\~{a}o Paulo
model.

However in more detail, there are very interesting differences between our calculations and the S\~{a}o Paulo model.  These
differences are likely due to additional dynamics, that is included in our calculations but that is neglected in the S\~{a}o
Paulo model. For the asymmetric systems $^{12}$C+$^{24}$O or $^{16}$O+$^{24}$O we predict an order of magnitude larger
cross section than for the S\~{a}o Paulo model. This is likely due to the dynamical effects that change the nuclear
densities during the collision process caused by the rearrangement of the single-particle wavefunctions.
This enhancement of fusion cross sections of very neutron rich nuclei can be tested in the laboratory with
radioactive beams.

Finally, fusion in accreting neutron stars may take place in the inner crust where there is a background gas of
neutrons.  This background gas could impact fusion cross sections.  To investigate this we are performing TDHF
simulations where the initial conditions involve two nuclei, appropriately boosted towards each other, and some unbound
neutrons, that are uniformly distributed in the simulation volume.  We will report these results in a later
publication.

\section*{Acknowledgments}
We thank Ed Brown, Hendrik Schatz and Witek Nazarewicz for helpful discussions and acknowledge the hospitality of the
Physics Division of Oak Ridge National Laboratory where this work was started.  This work was supported in part by DOE
grant Nos. DE-FG02-87ER40365 and DE-FG02-96ER40963.

\end{document}